\documentclass{llncs}

\usepackage{amsmath}
\usepackage{amssymb, graphicx}

\usepackage[latin1]{inputenc}
\usepackage[english]{babel}

\usepackage{gastex}

\usepackage{mathrsfs}

\newcommand{\hf}{\hfill}

\newcommand{\A}{ \mathbf{A}}
\newcommand{\C}{ \mathbf{C}}
\newcommand{\G}{ \mathbf{G}}
\newcommand{\T}{ \mathbf{T}}

\newcommand{\st}{\mathop{stem}}
\newcommand{\lo}{\mathop{loop}}
\newcommand{\supp}{\mathop{supp}}

\newcommand{\debcor}{\begin{corollary} }
\newcommand{\debexm}{\begin{example} }
\newcommand{\debdem}{\begin{proof} }
\newcommand{\debctr}{\begin{center} }
\newcommand{\debthm}{\begin{theorem} }
\newcommand{\debarr}{\begin{array} }
\newcommand{\debitem}{\begin{itemize} }
\newcommand{\debeqn}{\begin{eqnarray}}   
\newcommand{\debeqno}{\begin{eqnarray*}} 

\newcommand{\fincor}{\end{corollary} }
\newcommand{\finexm}{\end{example} }
\newcommand{\findem}{\end{proof} }
\newcommand{\finctr}{\end{center} }
\newcommand{\finthm}{\end{theorem} }
\newcommand{\finarr}{\end{array} }
\newcommand{\finitem}{\end{itemize} }
\newcommand{\fineqn}{\end{eqnarray}}   
\newcommand{\fineqno}{\end{eqnarray*}} 

\newcommand{\inc}{\subseteq}

\newcommand{\w}{\omega}

\renewcommand{\S}{\Sigma}

\newcommand{\ov}{\overline}

\def\by#1{\mathop{{\hbox{\setbox0=\hbox{$\scriptstyle{#1\quad}$}{$%
\mathrel{\mathop{\setbox1=\hbox to \wd0{\rightarrowfill}\ht1=3pt\dp1=-2pt\box1}\limits^{#1}}%
$}}}}}

\newcommand{\ra}{\rightarrow}

\newcommand{\m} [1]{{\mathcal{#1}}}

\title{Attenuation Regulation \\ as a Term Rewriting System
\thanks{The support of CNRS-RAS cooperation agreement 19122 \textsc{Evolver} is gratefully acknowledged.}}
 \author{Eugene Asarin\inst{1}\and Thierry Cachat\inst{1}\and Alexander Seliverstov\inst{2}\and\\
   Tayssir Touili\inst{1}\and Vassily Lyubetsky\inst{2}}

\institute{ LIAFA, CNRS and University Paris Diderot, \email{asarin,txc,touili@liafa.jussieu.fr} \and
  IITP, Russian Academy of Science, \email{slvstv,lyubetsk@iitp.ru}}

\begin{document}

\maketitle

\begin{abstract}
 The  classical attenuation regulation of gene expression  in bacteria is considered. We propose
to represent  the secondary RNA structure in the leader region of a gene or an operon  by a term,
and we give a probabilistic term rewriting system modeling the whole process of such a regulation.
\end{abstract}

\section{Introduction}

Modeling the mechanisms of regulation of gene expression, allowing  prediction of quantitative
characteristics of this expression (such as estimation of the level of expression  and
concentration of the substrate) is an important research challenge. In a previous work
\cite{Lyu06,Lyu07}, a model of one particular kind of regulation, the classical attenuation regulation,
has been suggested. In that model, the evolution of the secondary RNA structure  in the leader
region of a gene, and the progress of the ribosome and the polymerase along the RNA/DNA strands,
are represented  by a very special, elaborated in detail, Markov chain.  In this chain the
transition probability corresponding to the progress of the ribosome  depends on a ``control
variable'' --- the concentration of charged tRNA molecules in the cell. All the other
probabilities do not depend on the control variable, they can be determined from energy-based
considerations. Termination and antitermination (of gene expression) correspond to particular
random events in the Markov chain. In   \cite{Lyu06}, a Monte-Carlo simulation of this Markov
chain led to biologically realistic dependence of termination probability from the control
variable. Due to a large size and a complex structure of the Markov chain, its simulation is a
heavy computational task, but it was successfully solved, and a software tool called
\textsc{Rnamodel} simulates one trajectory in fractions of a second \cite{Lyu06,rnamodel}.
However, the approach based on the direct description of the Markov chain and its simulation has
some limitations, especially for a theoretical analysis. Biologically, it would be nice to have a
more structured and compact representation of the Markov chain and its instantaneous probability
distributions over all states at every instant, or only for sufficiently large time, or only
probabilities of the two biologically important events --- termination and antitermination.

Note that the problem of modeling the classical attenuation regulation, as stated in \cite{Lyu06}
and in the current article, is related to the representation of the transient behavior of the
secondary structure on a sliding window on the RNA strand between the ribosome and the polymerase
(see below for details). This differs from the kinetics of the secondary RNA structure on a fixed
nucleotide sequence for unlimited time, i.e. unlimited number of steps, investigated in many
papers. The  structure that appears after a large amount of time is called equilibrium secondary
RNA structure, it corresponds to a minimum of energy, see e.g. \cite{Zuker03,flamm}.
The tool
\textsc{Rnamodel} has also the function of determining this equilibrium  structure and
its
energy  as a special part of the full model in \cite{Lyu06}. However, real structures that appear
on the RNA strand during the regulation process are far from the equilibrium and their energies
are far from minimal.

In this article we discover a regular internal structure of the Markov chain describing the
classical attenuation regulation. We show that it can be represented as a probabilistic term
rewriting system for a particular type of terms. The set of rewriting rules can be large, but
all of them are generated by a small set of (five) metarules. In fact we give the full description
of the metarules and explain how to generate all the rules for the case  of classical attenuation
regulation.

Potential benefits of such a representation are  multiple:
\begin{itemize}
  \item easier and more precise modeling of  regulation mechanisms depending on the dynamics of
the secondary structure;
  \item compact description of such mechanisms, perhaps in dedicated languages, and hence a better
biological understanding of regulation processes;
  \item convenient representation of secondary structures by terms;
  \item specific analysis and simulation methods for rewriting systems.
\end{itemize}

This article is structured as follows. In section \ref{sec:bio} we describe shortly the biological
phenomenon that we want to model: the mechanism of classical attenuation regulation (CAR).
 In  section \ref{sec:terms} we introduce a class of terms and probabilistic  term rewriting
systems.  In section \ref{sec:metamodel} we represent a qualitative metamodel of the  biological
mechanism of CAR by a term rewriting system.  In section \ref{sec:model}
we refine the previous system and decorate its transitions with rates, thus obtaining a
representation of the Markov chain by a probabilistic term rewriting system. In section
\ref{sec:exp} we show some simulation results.  In section \ref{sec:related} we discuss some related work on term rewriting and its applications.
 In section \ref{sec:conclusion} we conclude with a
discussion of perspectives of the rewriting approach to modeling the mechanisms involving RNA
secondary structures, especially regulation.

\section{Classical attenuation regulation}\label{sec:bio}
To begin with, we recall some well-known biological facts about the biological phenomenon playing the central
role in this article.

\begin{figure}
  \includegraphics[width=0.9\textwidth]{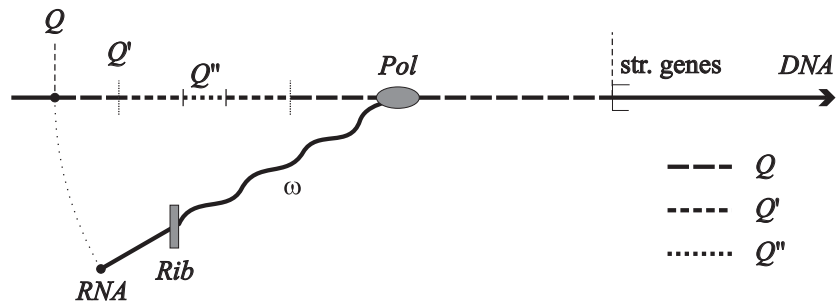}\\
  \caption{Classical attenuation regulation. The RNA polymerase \emph{Pol} transcribes the
regulatory region $Q$, the ribosome \emph{Rib} translates the leader peptide gene $Q'$. The
movement of \emph{Rib} on regulatory codons $Q''$ is controlled by the concentration of charged
tRNA. The secondary RNA structure $\omega$ between \emph{Rib} and \emph{Pol} brakes \emph{Pol}
and pushes it  off the chain. If \emph{Pol} reaches the structural genes, then they are expressed,
i.e. transcribed and then translated.
Note that in both the DNA and the RNA, we use  $Q, Q'$ and $Q''$ to denote    the regulatory region,  the leader peptide gene, and the
 regulatory codons, respectively.} \label{fig:car}
\end{figure}

The expression of a group of structural genes (that is synthesis of the corresponding proteins,
which are ferments for a chemical reaction) can be regulated by a sequence of nucleotides placed
on the DNA upstream  inside the so called \emph{leader region} of the genes \cite{SB91}.
This subsequence of the leader region is called the \emph{regulatory region}. In this article we
deal with one particular  type of regulation, \emph{classical attenuation regulation (CAR)} in
bacteria. This regulation mechanism  concerns structural genes (groups of genes --- operons) that
produce proteins which catalyze the synthesis of amino acids.  The classical attenuation allows to
activate such an operon when the cell contains a small concentration of the amino acid, to
deactivate the operon whenever this concentration increases, and to do it fast. The mechanism of
CAR involves several actors: the regulatory region on the DNA, its copy on the RNA, the ribosome,
and a ferment called RNA polymerase (see Fig.\ref{fig:car}).

For  structural genes to be expressed two concurrent processes should succeed: the regulatory
region $Q$  should be transcribed creating an RNA by RNA polymerase. At the same time the ribosome
should be bound to the very beginning of the freshly created segment $Q'$ (called the \emph{leader
peptide gene}) in the regulatory region $Q$ on the RNA and starts translation  of this leader
peptide gene to an auxiliary protein.  The essential part of the  regulation process takes place
when the ribosome moves on $Q'$ on the RNA  and the polymerase moves somewhere downstream of the
ribosome on $Q$ on the DNA.

The ribosome moves ``rightwards'' (formally speaking, in the direction from the $5'$ to the $3'$
end) on a segment $Q'$ of the sequence $Q$. Its speed is constant except on a subsequence $Q''$
(\emph{regulatory codons}) where it depends directly on the concentration of the amino acid (via
charged tRNA concentration). To the right of the ribosome and independently of it, the polymerase
moves rightwards on $Q$. Between the ribosome and the polymerase a secondary structure $\omega$
is formed on the RNA. This structure consists in pairing of some nucleotides, and it changes very
fast. An important effect of the secondary structure $\omega$ consists in slowing down the
movement of the polymerase. There are two possible scenarios:
\begin{itemize}
  \item When $\omega$ is strong enough, its ``braking'' action on the polymerase increases, and
moreover, the polymerase can slip off the DNA (this can only happen on so-called \emph{T-rich
sequence}, where the connection of the polymerase and the DNA weakens). Such an event is called
\emph{termination}, and in this case the structural genes are not expressed: the transcription of
the regulatory region is aborted, the structural genes  are not transcribed and therefore not
translated.
  \item  Another possibility is that the ribosome moves fast enough to weaken or partly destroy
most of the structure $\omega$. In this case the polymerase safely traverses the T-rich sequence,
and arrives to the end of the leader region $Q$. Next, the polymerase enters the structural genes,
and their transcription, followed by translation are unavoidable. This event is called
\emph{antitermination} and in this case the structural genes are expressed.
\end{itemize}

In the rest of this article we build a qualitative and a quantitative models of the regulation
process described above.

\section{Terms and rewriting systems}\label{sec:terms}

\subsection{Unranked unordered terms}

Let $\Sigma$ be a finite set of function symbols and $\m{X}$ an enumerable set of {\em variables}
(standing for sets of terms).
The set $T_{\Sigma}[\mathcal{X}]$ of terms over $\Sigma$ and $\mathcal{X}$ is  the smallest set
that satisfies:
\begin{itemize}
  \item $\Sigma\subseteq T_{\Sigma}[\mathcal{X}]$,
  \item $\{f(x)\mid f\in\Sigma\land x\in\mathcal{X}\}\subseteq T_{\Sigma}[\mathcal{X}]$,
  \item if $f\in\Sigma$ and 
$s\inc T_{\Sigma}[\mathcal{X}]$ is a {\em set} of terms, then $f(s)$ is in $T_{\Sigma}[\mathcal{X}]$.
\end{itemize}

By definition we also put $f(\emptyset)=f$ for $f\in\Sigma$. For convenience we write $f(g,h(e))$
instead of $f(\{g,h(\{e\})\})$. However one should remember that the coma-separated terms are unordered.

\debexm
	Let $\S=\{e,f,g,h\}$ and $\m{X}=\{x,y,z,\dots\}$, then  the followings are terms in
	$T_{\Sigma}[\mathcal{X}]$:
		$f(g,h(e))$, $f(f(x))$ and $e(g,f)$.
\finexm

Note that we consider function symbols of variable arity. $T_{\Sigma}$  stands for $T_{\Sigma}[\emptyset]$.
Terms in $T_{\Sigma}$ are called {\em ground terms}. Variables are used only to define substitution and
rewriting rules. The ``real'' terms are ground terms.
A {\em substitution} $\sigma$ is a mapping from $\m{X}$ to $2^{T_{\Sigma}[\mathcal{X}]}$,
written as $\sigma=\{x_1\ra T_1,\ldots, x_n\ra T_n\}$, where $T_i$, $1\le i\le n$, is a finite set of  terms that
substitutes the variable  $x_i$. The term obtained by applying the substitution $\sigma$ to a term $t$ is
written $t\sigma$. We call it an {\em instance} of $t$.

Let $R$ be a rule of the form $l\ra r$, where $l$ and $r$ are terms in $T_{\Sigma}[\mathcal{X}]$.
For ground terms $t, t'$ we write $t\ra_R t'$ if there exists a substitution $\sigma$ such that $t'$ can be obtained from
$t$ by replacing an occurrence of  the subterm
$l\sigma$ by $r\sigma$.  $\ra_R$ defines a relation between ground terms. Let  $\ra_R^*$ be the reflexive transitive closure of $\ra_R$.

\debexm
	Let $R=l\ra r$ with $l=f(x,e)$, $r=f(g(x),e)$ and $t=e(f(h,e))$, then $t\ra_R t'$ where
	$t'=e(f(g(h),e))$.
\finexm

A \emph{term rewriting system (TRS)}  is a finite set of rules of the form $l\ra r$. Given a TRS $\mathcal{R}$
and a set of terms $I\subset T_{\Sigma}$, the language $\mathcal{R}^*(I)$ is defined as the set of all ground
terms that can be obtained from the terms in $I$ by applying a finite number of times the rules from  $\mathcal{R}$, i.e.,
$\mathcal{R}^*(I)=\{t\in  T_{\Sigma}\mid \exists t'\in I, t'\ra_{\mathcal{R}}^* t\}$.

\debexm
	Let $\mathcal{R}=\{f(x)\ra g(f(x))\}$  and $I=\{f(e,h)\}$, then
    $$\mathcal{R}^*(I)=\{g^n(f(e,h))\mid n\in \bbbn\}.$$
\finexm

\subsection{Probabilistic Term Rewriting Systems}

A {\em Continuous Time Markov Chain} 
is a pair $(S,\rho)$, where $S$ is a finite or enumerable set of states and $\rho:S\times S\to [0,\infty)$ is the rate matrix.
For $s,s'\in S$, $\rho(s,s')>0$ means that there is a transition between states $s$ and $s'$, and that the probability
for moving from $s$ to $s'$ within $t$ time units is equal to $1-e^{-\rho(s,s')\cdot t}$.
If a state $s$ has more than one outgoing transition (i.e., if there exist more than one state $s'$ for which $\rho(s,s')>0$)
there exists a {\em race} between these transitions and the probability
for moving from $s$ to $s'$ within $t$ time units is equal to $\frac{\rho(s,s')}{E(s)}\big(1-e^{-E(s)\cdot t}\big)$, where
$E(s)=\sum\limits_{s'\in S} \rho(s,s')$.

A (continuous time) {\em Probabilistic term rewriting system (PTRS)} over $\Sigma\cup \m{X}$ is a (finite) set of rules of the form $l\by{\Lambda}r$,
 where $l$ and $r$ are terms in $T_{\Sigma}[\mathcal{X}]$, and $\Lambda\in(0,\infty)$ is a rate.

A PTRS $\m{R}$ over $\Sigma\cup \m{X}$ defines a continuous time Markov chain on ground terms $M=(T_\Sigma,\rho)$, where $\rho(t,t')=\Lambda$ iff there exists a rule  $l\by{\Lambda}r \in\m{R}$ such that $t\ra_R t'$, where $R$ is the ``non probabilistic'' rule $l\ra r$.

\begin{remark}
  If there are several rules (or several instances of the same rule) that lead from $t$ to $t'$, then $\rho(t,t')=\sum\Lambda$, where the sum is taken over all such rules or instances.
\end{remark}

\section{Metamodel}\label{sec:metamodel}
We want to model the phenomenon of the classical attenuation regulation described in section \ref{sec:bio}.

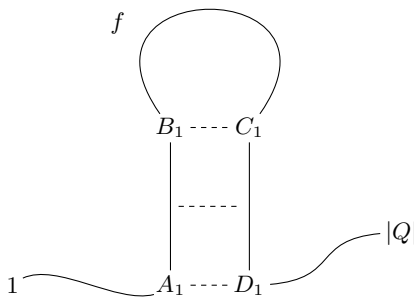
\begin{figure}
\begin{center}
{ \unitlength=0.7mm
\begin{picture}(80,60)(7,5)
	\gasset{Nadjust=wh,Nadjustdist=1,Nframe=n}

	\node(1)(10,10){$1$}
	\node(f1)(40,10){$A_1$}
	\node(f2)(40,40){$B_1$}
	\node(f3)(55,40){$C_1$}
	\node(f4)(55,10){$D_1$}
	\node(e)(84,20){$|Q|$}
\node(name)(30,60){$f$}

	\gasset{AHnb=0}
	\drawbpedge(1,45,10,f1,-135,10){$$}
	\drawedge(f1,f2){$$}
	\drawbcedge(f2,15,70,f3,80,70){$$}
	\drawedge(f3,f4){$$}
	\drawbpedge(f4,0,20,e,-180,20){$$}

	\gasset{dash={1 1}0}	
	\drawedge(f1,f4){$$}
	\drawedge(f2,f3){$$}

	\gasset{Nadjustdist=1.5}
	\node(fa)(40,25){$$}
	\node(fb)(55,25){$$}
	\drawedge(fa,fb){$$}
\end{picture} }
\end{center}
\caption{One hypohelix $f$. }\label{fig:hypohelix}
\end{figure}

We suppose that a {\em regulatory region} $Q$ (see Fig. \ref{fig:car})  is given and fixed in the sequel, it  is a sequence (word)
$Q\in\{\A,\C,\G,\T\}^*$,  the letters of this alphabet are called \emph{nucleotides}.
We denote by $|x|$ the length of any word $x$ and $x_i$ the $i$th letter of $x$, so
$x=x_1 x_2 \dots x_{|x|}$.
The sequence $Q$  can be  folded\footnote{only on its ``active'' part called \emph{window}, as we  will see below} in a way that  some nucleotides of $Q$ are paired: $\A$ with $\T$ and $\C$ with $\G$.
The complement of a nucleotide is written using a bar:
$\A=\ov{\T}$, $\T=\ov{\A}$, $\C=\ov{\G}$, $\G=\ov{\C}$.
  We look in $Q$ for subwords (``stems'') of the form
\debeqn
        && Q_{A}Q_{A+1}\dots Q_{B}\ \ \mbox{ and }\ \ Q_{C}Q_{C+1}\dots Q_{D}
           \ \ \mbox{ such that }\ \ \nonumber\\
	&& B-A=D-C,\ A+3\leq B,\ B+3\leq  C\label{eq:hypohelix}\\
	&& Q_{A}=\ov{Q_{D}},\ Q_{A+1}=\ov{Q_{D-1}}, \dots\ Q_{B}=\ov{Q_{C}}\ . \nonumber
\fineqn
Any pair of such stems forms a \emph{hypohelix} (see Figure \ref{fig:hypohelix}, where the labels $A_i,B_i,C_i$ and $D_i$ are positions in the word $Q$).
\\
We describe a hypohelix $f$ by a tuple of its stems' extremities  $f=(A,B,C,D)$, and we introduce the following notations:
\begin{eqnarray*}
  \st(f) = [A,B] \cup [C,D],\
  \lo(f) = [B+1,C-1], \
  \supp(f) = [A,D].
\end{eqnarray*}

There is a {\em ribosome} at some position on  $Q'$ and an {\em RNA polymerase} somewhere to the right of it.
Both move to the right, in one step the ribosome moves by three successive
nucleotides and the polymerase by one nucleotide. The \emph{window}  $w=(R,P)$  represents the segment of RNA
from the first position $R$ after the end  of the ribosome to the last position  $P$ before the beginning of
the polymerase. In fact the folding of the RNA sequence $Q$ can only happen within the current window, i.e.
between positions $R$ and $P$.  When the ribosome advances to the right, it can destroy the leftmost
hypohelix of a  current configuration, because it consumes the first three letters of the window.
On the other hand any polymerase move adds one new letter to the window.

Formally a window has the form  $w=(R,P)$ with $R,P\in \bbbn$.
 The following constraints should be satisfied:
\begin{equation}
  13\le R \le P  \le |Q|  \label {eq:window}
\end{equation}

Thus, the window is moving and changing its length.

Let $W=\{w=(R,P)\mid \mbox{ conditions (\ref{eq:window}) are satisfied} \}$ be the alphabet of all windows.
 We define
\begin{eqnarray*}
  \st(w) = \emptyset,\
  \lo(w) = [R,P],\
  \supp(w) = [R,P]\ .
\end{eqnarray*}

We will write terms over the alphabet $\S$ of all hypohelices and all windows:
\debeqno
	\Sigma= H  \cup W \mbox{ where }
	H=\{ f =(A,B,C,D)\mid \textrm{conditions (\ref{eq:hypohelix}) are satisfied} \}.
\fineqno
We consider only terms of the form $w(\dots)$ for some $w\in W$ (rooted by some window $w$).
According to the conditions that we will define next, a symbol $f =(A,B,C,D)$ can appear
in a term $w(\dots)$ only if $R\leq A$ and $D\leq P$, where $w= (R,P)$.

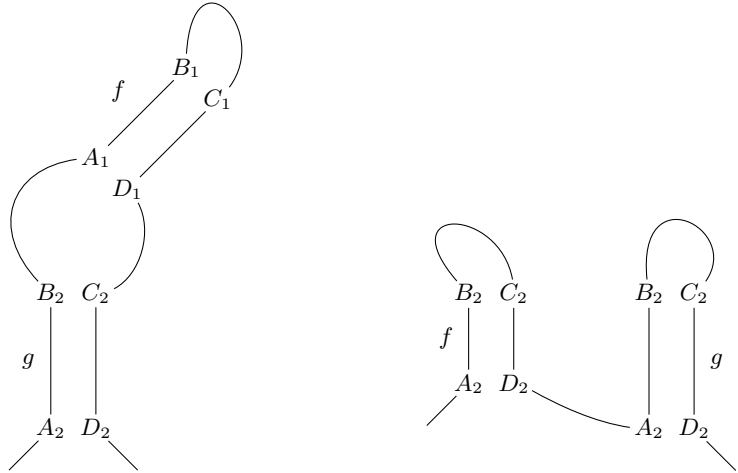
\begin{figure}
\begin{center}
\hf
{ \unitlength=0.6mm
\begin{picture}(60,110)(25,-62)
	\gasset{Nadjust=wh,Nadjustdist=1,Nframe=n}

	\node(f1)(50,10){$A_1$}
	\node(f2)(70,30){$B_1$}
	\node(f3)(77,23){$C_1$}
	\node(f4)(57,3){$D_1$}
\node(namef)(55,25){$f$}

	\gasset{AHnb=0}
	\drawedge(f1,f2){$$}
	\drawbpedge(f2,90,30,f3,40,20){$$}
	\drawedge(f3,f4){$$}

	\node(1)(30,-60){$$}
	\node(g1)(40,-50){$A_2$}
	\node(g2)(40,-20){$B_2$}
	\node(g3)(50,-20){$C_2$}
	\node(g4)(50,-50){$D_2$}
	\node(e)(60,-60){$$}
\node(nameg)(35,-35){$g$}

	\drawedge(1,g1){$$}
	\drawedge(g1,g2){$$}
	\drawedge(g3,g4){$$}
	\drawedge(g4,e){$$}

	\drawbpedge(g2,140,20,f1,-180,20){$$}
	\drawbpedge(g3,0,10,f4,-40,10){$$}
\end{picture} }
\hf
{ \unitlength=0.6mm
\begin{picture}(80,60)(-5,-2)
	\gasset{Nadjust=wh,Nadjustdist=1,Nframe=n}

	\gasset{AHnb=0}
	\node(1)(0,10){$$}
	\node(f1)(10,20){$A_2$}
	\node(f2)(10,40){$B_2$}
	\node(f3)(20,40){$C_2$}
	\node(f4)(20,20){$D_2$}
\node(namef)(5,30){$f$}

	\drawedge(1,f1){$$}
	\drawedge(f1,f2){$$}
	\drawbpedge(f2,135,30,f3,90,20){$$}
	\drawedge(f3,f4){$$}

	\node(g1)(50,10){$A_2$}
	\node(g2)(50,40){$B_2$}
	\node(g3)(60,40){$C_2$}
	\node(g4)(60,10){$D_2$}
	\node(e)(70,0){$$}
\node(nameg)(65,25){$g$}

	\drawedge(g1,g2){$$}
	\drawedge(g3,g4){$$}
	\drawedge(g4,e){$$}
	\drawbpedge(g2,100,30,g3,40,20){$$}

	\drawbpedge(f4,140,20,g1,-180,20){$$}
\end{picture} }
\hf
\end{center}
\caption{Relative positions of two hypohelices $f$ and $g$: $f\prec g$ and $f\bowtie g$.
Here $f=(A_1,B_1,C_1,D_1)$ and $g=(A_2,B_2,C_2,D_2)$. On the left $B_2<A_1$ and $D_1<C_2$,
on the right $D_2<A_2$.
}
\label{fig:positions}
\end{figure}

We say that a hypohelix $f$ is {\em embedded} in   $g$ (which can be a hypohelix or a window), written $f\prec g$, if $\supp(f)\subseteq \lo(g)$.
Two hypohelices $f$ and $g$ are \emph{disjoint}, written $f\bowtie g$, if $\supp(f)\cap\supp(g)=\emptyset$.
We call $f$ and $g$ \emph{unknotted} if either one of them is embedded in the other or they are disjoint.
We say that $g=(A_2,B_2,C_2,D_2)$ is an \emph{extension} of $f=(A_1,B_1,C_1,D_1)$, denoted $f\sqsubseteq g$, if $[A_1,B_1]\subseteq [A_2,B_2]$ and $B_2-B_1=C_1-C_2$, hence
$[C_1,D_1]\subseteq [C_2,D_2]$, and the pairing in $g$ is an extension of that in $f$.
See Figure \ref{fig:positions}.

We call a term $t$ over $\S$ \emph{well-formed} if it satisfies the following conditions:
\begin{description}
      \item[(compatibility)] any  $f$ and $g$ appearing in $t$  are unknotted, in particular any $f$ can appear at most once,
      \item[(ordering)] if $f$ and $g$ occur in $t$, then
	  $f\prec g$  iff $f$ is in the scope of $g$.
\end{description}

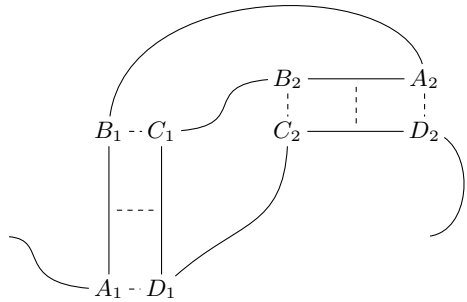
\begin{figure}
\begin{center}
{ \unitlength=0.7mm
\begin{picture}(95,60)(15,5)
	\gasset{Nadjust=wh,Nadjustdist=1,Nframe=n}

	\node(1)(20,20){$$} 	\node(f1)(40,10){$A_1$} 	\node(f2)(40,40){$B_1$} 	\node(f3)(50,40){$C_1$}
	\node(f4)(50,10){$D_1$} 	\node(e)(100,20){$$}
	\node(g1)(100,50){$A_2$} 	\node(g2)(74,50){$B_2$} 	\node(g3)(74,40){$C_2$}
	\node(g4)(100,40){$D_2$}   	\node(fa)(40,25){$$} 	\node(fb)(55,25){$$}

	\gasset{AHnb=0} 	\drawbpedge(1,0,10,f1,180,20){$$} 	\drawedge(f1,f2){$$}
	\drawbpedge(f2,90,30,g1,90,20){$$} 	\drawedge(g1,g2){$$} 	 \drawbpedge(g2,180,20,f3,0,20){$$}
	\drawedge(f3,f4){$$} 	\drawbpedge(f4,45,20,g3,-90,20){$$} 	\drawedge(g3,g4){$$}
	\drawbpedge(g4,0,10,e,0,10){$$}
	\gasset{dash={1 1}0}	 	\drawedge(f1,f4){$$} 	\drawedge(f2,f3){$$}
	\drawedge(g1,g4){$$} 	\drawedge(g2,g3){$$}
	\gasset{Nadjustdist=1.5}
	\node(fa)(40,25){$$} \node(fb)(50,25){$$} \drawedge(fa,fb){$$}
	\node(fa)(87,40){$$} \node(fb)(87,50){$$} \drawedge(fa,fb){$$}
\end{picture} } \\
\end{center}
\caption{Pseudo-knot: $A_1<B_1 <A_2 <B_2 <C_1 <D_1 <C_2 <D_2$. Such configurations are not allowed in our model. }
\label{fig:pseudoknot}
\end{figure}

The combination of two hypohelices  in Figure \ref{fig:pseudoknot} is biologically feasible, but according to our rules
these hypohelices are incompatible. We believe that this restriction (crucial for representation by terms) does not undermine significantly the accuracy of the model.

Notice, that a well-formed term of the form $w(\dots)$ (rooted by some window $w$) contains only hypohelices from
$$\Sigma_w= \{ f\in H\mid f\prec w\}.
$$
This simple observation greatly simplifies the simulation process.

In \cite{Lyu06} an additional \emph{maximality} condition is imposed. Using the terminology of this article,
it requires that  no hypohelix $f$ in $t$ can be replaced by its proper extension without creating an
overlapping. Here we do not impose this restriction.

Each well-formed term represents a possible secondary RNA structure in a window in  $Q$: the set of hypohelices
that are present in this window. It could be possible to allow knotted hypohelices, and hypohelices of length
less than 3, but here we do not consider them.

We extend the definitions of $\bowtie$ and $\prec$:
let $f$ be a term and $\vec{c}$ a set of terms,
\debeqno
	\vec{c}\bowtie f \mbox{ iff }  \forall g\in \vec{c}\,  (g\bowtie f),\\
	\vec{c}\prec f    \mbox{ iff }  \forall g\in \vec{c}\,  (g\prec f).
\fineqno
In the former case we say that $f$ and $\vec{c}$ are \emph{disjoint}, in the latter that $\vec{c}$ is embedded into $f$.

We start from a sequence $Q$  without any pairing of nucleotides, this structure is described
by a term $w()$ --- ``an empty window'', where $w=(13,13)$.
Our aim is to represent the evolution of the secondary structure in the window, as well as the progress of
the ribosome and the polymerase, through rewriting terms
starting from $w()$. Our rewriting system will generate only well-formed terms.

On the whole, there are five rewriting {\em Meta}-rules:
\debitem
      \item  Binding and decomposition of a hypohelix $f$:
	\debeqn
	   \big(\omega=g(\vec{c},\vec{d})\big)\ \longleftrightarrow\ \big(\omega'=g(\vec{c},f(\vec{d}))\big)\quad
	   \mbox{ with } \vec{c}\bowtie f,\ \vec{d}\prec f,\ f \prec g,   \label{eqn-a}
	\fineqn
	where $\vec{c}$ and $\vec{d}$ are sequences of terms. The {\em concrete} rewriting
	rules --- and their rates --- depend on $\vec{c}$ and $\vec{d}$, as explained below.
      \item Extension and reduction of a hypohelix
	\debeqn
		\big(\omega=f\big) \longleftrightarrow \big(\omega'=g\big) \quad
	   	\mbox{ with } f \sqsubseteq g.  \label{eqn-b}
	\fineqn

      \item The window movement can be described by the following rules, where $w=(R,P)$:
\finitem

\debeqn
(R,P)(\omega) &\longrightarrow &  (R+3,P)(\omega')\ , \label{eqn-c} \\
(R,P)(\omega) &\longrightarrow & (R,P+1)(\omega)\ , \label{eqn-d} \\
w(\omega) 	&\longrightarrow& \bot\ .  			 \label{eqn-e}
\fineqn

\medskip
In the last rule, $\bot$ is a special symbol denoting termination.
Rules (\ref{eqn-c}) describe the movement of the ribosome. In these rules,  $\omega'$ is obtained from $\omega$ by removing only the possible symbol
that is incompatible with the new window $(R+3,P)$, or replacing it by a ``shorter'' hypohelix.
Indeed, if the leftmost hypohelix in $\omega$ starts at a position between $R$ and $R+3$, then the movement of the ribosome
by three  positions to the right will destroy this hypohelix. More formally, if  $\omega\prec (R+3,P)$,
 then $\omega'=\omega$. Otherwise the ribosome destroys  the leftmost hypohelix.
 In this case,  there is a single symbol $f$ in $\omega$ such that $f\not\prec (R+3,P)$.
Suppose the subterm rooted by $f$ is $f(\vec{c})$. Then, $\omega'$ is obtained by  replacing in  $\omega$ $f(\vec{c})$ by  either $f'(\vec{c})$ or $\vec{c}$,
depending on the size of $f$,  where $f'\sqsubseteq f$.

Rules \ref{eqn-d} describe the movement of the  polymerase. Note that if the   polymerase reaches a position $P+1$ where the structural genes are expressed, then
we reach antitermination
and the gene is expressed.

\section{Quantitative model} \label{sec:model}

Now, we introduce the rates of the five rewriting rules.

Let $h(f_1(*),\dots,f_n(*))$ be a term. Then the {\em free loop length} of the hypohelix $h$ in this term is
$$
l_h=|\lo(h)| - \sum_{i=1}^n |\supp(f_i)|\ .
$$
This numeric characteristic corresponds to the number of nucleotides in the loop of the hypohelix $h$ that
do not participate in inner hypohelices.

In order to define the rate, we have to consider the concrete rule corresponding
to the Metarule (\ref{eqn-a}).
For any $f,g,\vec{c}=c_1(x_1),\dots,c_m(x_m)$ and $\vec{d}=d_1(y_1),\dots,d_n(y_n)$ such that
$\vec{c}\bowtie f,\ \vec{d}\prec f,\ f \prec g$ there is a concrete rule
	\debeqn
	 &  \big(\omega=g(c_1(x_1),\dots,c_m(x_m), & d_1(y_1),\dots,d_n(y_n))\big)\ \nonumber \\
\longleftrightarrow\ & \big(\omega'=g(c_1(x_1),\dots,c_m(x_m), & f(d_1(y_1),\dots,d_n(y_n)))\big) \label{eqn-a'}
	\fineqn
Recall that the subterms are unordered.
Similarly the concrete rule corresponding to (\ref{eqn-b}) is
	\debeqn
	&  \big(\omega=a(c_1(x_1),\dots,c_m(x_m), & f(d_1(y_1),\dots,d_n(y_n)))\big)\ \nonumber \\
\longleftrightarrow\ & \big(\omega'=a(c_1(x_1),\dots,c_m(x_m), & g(d_1(y_1),\dots,d_n(y_n)))\big) \label{eqn-b'}
	\fineqn
Note that this transformation can change the free loop length of the hypohelix $a$.
The rate of the rules (\ref{eqn-a'}-\ref{eqn-b'}) is denoted  $K(\w \ra \w')$, given by
\debeqn
	K(\w \ra \w')= \kappa\cdot\exp\left(\frac{1}{2}\left(E(\w)-E(\w')\right)\right),
\fineqn
where the energy  $E(\w)= G_{hel}(\w)+ G_{loop}(\w)$, $\kappa$ is a
 parameter
--- usually $\kappa=10^3$ --- and
\debeqn
	G_{hel}(\w)=\frac{1}{RT}\cdot \sum_h E_{h} & \mbox{ and } &
	G_{loop}(\w)= \sum_h 1.77\cdot  \ln(l_h +1)+ B\ ,
\fineqn
and $h$ varies over all hypohelices  from $\w$.
 $E_{h}$ represents the total stacking energy along the hypohelix $h$. It is the sum of stacking bond energies of the adjacent base pairs of $h$.
 $B$ can take three different values depending on the three possible types of the loop of the
hypohelix $g$: terminal loop, single-strand bulge and double-strand bulge.

A {\em codon} is a triple of successive nucleotides. For a sequence $Q'$, each codon
is fixed to be either regulatory  or non-regulatory.
Analogously, each nucleotide in $Q$ is fixed to be either non T-rich or T-rich \cite{Lyu06}.
Let  $s_0$ be the ``radius'' of a ribosome --- distance from P-site to the end of the
ribosome --- usually $s_0=12$, and let  $s_1$ be  the ``radius'' of a polymerase --- distance from the $5'$ end of a polymerase
to its transcription center --- usually $s_1=9$.
The rate  of the  rule (\ref{eqn-c}) is denoted $\lambda_{rib}$ and
is constant when $R-s_0$ is a position of a non-regulatory codon,
and otherwise $\lambda_{rib}$ depends on an external parameter $c$ ---
the concentration of charged tRNA \cite{SB91}.
The rate of the rule  (\ref{eqn-d}) is denoted $\nu$ and depends on
secondary structure $\omega$ in the window.
The  rule (\ref{eqn-e})  applies only when $P+s_1$ is a position of a T-rich nucleotide
and its rate is denoted $\mu$.

In \cite{Lyu06} the rate of the rule (\ref{eqn-c}) was denoted $\lambda_{rib}$ and
\debeqn
	\lambda_{rib}(c)=\frac{45\,c}{1+c}\ .
\fineqn
The rate of the rule (\ref{eqn-d}) was denoted $\nu$ and
\debeqn\label{eq:rate-d}
	\nu=40-F(\omega)\ .
\fineqn
The rate of the rule (\ref{eqn-e}) was denoted $\mu$ and
\debeqn\label{eq:rate-e}
	\mu=\frac{1}{4}F(\omega)\ .
\fineqn
The function $F(\w)$ in (\ref{eq:rate-d}-\ref{eq:rate-e}) for $\omega=f_1(*),\dots, f_n(*)$ depends only on functional symbols
(hypohelices) $f_1,\dots, f_n$, and not on the structure of their arguments denoted by $*$.
More precisely $F(\w)= \max_i F(f_i)$, where
\debeqn \label{eq:force}
      F(f)=\frac{\delta\cdot  \exp\left(-\frac{r(f)}{r_0}\right)}{(L_2)^2\cdot (p(f)-p_0)^2+1}\ ,
\fineqn
with $p(f)\approx \frac{\pi}{|\supp(f)|}$, and $r(f)$  the ``free distance'' from $f$ to the
end $P$ of the window: for $f=(A,B,C,D)$ and $w=(R,P)$, we have
\debeqn
	r(f)=R-D-\sum_{i}|\supp(f_i)|\ .
\fineqn
Other symbols in equation (\ref{eq:force}) denote constants: $r_0=1, \delta=30, L_2=27.1, p_0=0.18$,
see \cite{Lyu06}.

Note  that the rates of the rules depend only on the local configuration as explained above
and not on the outside context. In particular it does not depend on instantiations of $x_1,\dots,x_m,y_1,\dots,y_n$.

\newcommand{\toshort}{\!\!\to\!\!}
\newcommand{\tostar}{\stackrel{*}{\to}}

\section{Simulation results} \label{sec:exp}
\begin{figure}[ht]
\begin{center}
\textsc
{
\input{EC_trp.seq}
}
\end{center}
\caption{A regulatory region for \emph{trpE} genes in  \emph{E.~coli}.}\label{fig:ecoli}
\end{figure}

\begin{figure}[ht]
\scalebox{0.75}{\parbox{1.33\textwidth}{
$
\input{EC_trp.term}
$
}}
\caption{\textbf{A simulation result:} one typical terminating trajectory for classical attenuation
regulation of \emph{trpE} genes in  \emph{E.~coli}. \textbf{Notations:} $\to$ means one rewriting;
$\tostar$ means  several similar rewritings; repeated window positions (e.g. repetitions of
$\langle 40,51 \rangle $)are replaced by a $\cdot$ symbol; $\bot$ means termination. There are 24
helices, denoted by letters from $a$ to $x$. }\label{fig:tool}
\end{figure}

We have adapted the simulator described in \cite{Lyu06} and available  at \cite{rnamodel} to  obtain
sequences of terms. As an example in Figure \ref{fig:tool} we give one (slightly shortened and simplified)
terminating trajectory of the regulation  process for the \emph{trpE} genes (responsible for the synthesis
of tryptophan) in \emph{E.~coli}. The regulatory region itself is presented in Figure \ref{fig:ecoli}.

\section{Related Work}\label{sec:related}
References  to the  literature on RNA  regulation mechanisms can be found in  \cite{Lyu06,Lyu07}.

Term rewriting systems have been used  in the so called {\em Regular Model Checking} framework
\cite{KMM+,BT02,ABMO:tree,julien-tacas}. They have been successfully applied to the analysis of parameterized systems
\cite{BT02,ABMO:tree,julien-tacas} and multithreaded programs \cite{BT02,BT03,Tou05}. However, in the regular model checking framework,
the rewriting rules are not probabilistic. This work constitutes  the first step towards the extension of the regular model checking framework with probabilistic rewriting rules. This would allow for example the analysis of  probabilistic parameterized systems and probabilistic multithreaded programs.

Rewriting systems have also been used in  articles \cite{RULE2005,RTA2003Chimie} to model chemical reactions. Compared to our work,
 the rewriting systems considered in \cite{RULE2005,RTA2003Chimie} are not probabilistic.
 Moreover, these works  consider the modeling of  chemical reactions whereas we consider modeling of RNA secondary structure.

Finally, probabilistic term rewriting systems have also been considered  in \cite{RTA2003Proba,ProbaRTA,kumar-rewriting}. But in these works, the
symbols are of fixed arities and the terms are ordered, whereas in our framework, the  symbols have arbitrary arities and the terms are not ordered.
Moreover, as far as we know, this is the first time that  probabilistic term rewriting systems are  used to model attenuation regulation.

\section{Conclusions and perspectives} \label{sec:conclusion}

We have established that the framework of probabilistic term rewriting systems provides compact and
structured description of detailed models of RNA regulation.

We intend to continue exploration of this framework. The most important task consists in the development
of adequate data structures and algorithms, as well as approximation and abstraction methods for
analysis of this kind of models. The next step would be a massive computational experimentation,
the biological interpretation of results and validation of results by real biological data.

\section*{Acknowledgments} The authors are thankful to Sergey Pirogov, Konstantin Gorbunov and Lev
Rubanov for a valuable discussion. Lev Rubanov has also provided assistance in use of the
\textsc{Rnamodel} tool. Oleg Zverkov has helped us in preparing computer graphics for this article.

\bibliographystyle{alpha}
\bibliography{termesRegul}

\newcommand{\etalchar}[1]{$^{#1}$}
\begin{thebibliography}{KMM{\etalchar{+}}01}

\bibitem[AJMd02]{ABMO:tree}
Parosh~Aziz Abdulla, Bengt Jonsson, Pritha Mahata, and Julien d'Orso.
\newblock Regular tree model checking.
\newblock In {\em CAV'02}, volume 2404 of {\em Lecture Notes in Computer
  Science}, pages 555--568, 2002.

\bibitem[ALdR05]{julien-tacas}
Parosh~Aziz Abdulla, Axel Legay, Julien d'Orso, and Ahmed Rezine.
\newblock Simulation-based iteration of tree transducers.
\newblock In {\em TACAS'05}, volume 3440 of {\em Lecture Notes in Computer
  Science}, pages 30--44, 2005.

\bibitem[BCC{\etalchar{+}}03]{RTA2003Chimie}
Olivier Bournez, Guy-Marie Côme, Valérie Conraud, Hélène Kirchner, and Liliana
  Ibanescu.
\newblock A rule-based approach for automated generation of kinetic chemical
  mechanisms.
\newblock In {\em RTA'03}, volume 2706 of {\em Lecture Notes in Computer
  Science}, pages 30--45. Springer, june 2003.

\bibitem[BH03]{RTA2003Proba}
Olivier Bournez and Mathieu Hoyrup.
\newblock Rewriting logic and probabilities.
\newblock In {\em RTA'03}, volume 2706 of {\em Lecture Notes in Computer
  Science}, pages 61--75. Springer, June 2003.

\bibitem[BIK06]{RULE2005}
Olivier Bournez, Liliana Ibanescu, and Hélène Kirchner.
\newblock From chemical rules to term rewriting.
\newblock In {\em 6th International Workshop on Rule-Based Programming}, volume
  147(1) of {\em ENTCS}, pages 113--134, 2006.

\bibitem[BK02]{ProbaRTA}
Olivier Bournez and Claude Kirchner.
\newblock Probabilistic rewrite strategies: Applications to {ELAN}.
\newblock In {\em RTA'02}, volume 2378 of {\em Lecture Notes in Computer
  Science}, pages 252--266. Springer-Verlag, July 2002.

\bibitem[BT02]{BT02}
Ahmed Bouajjani and Tayssir Touili.
\newblock Extrapolating tree transformations.
\newblock In {\em CAV'02}, volume 2404 of {\em Lecture Notes in Computer
  Science}, pages 539--554, 2002.

\bibitem[BT03]{BT03}
Ahmed Bouajjani and Tayssir Touili.
\newblock Reachability analysis of process rewrite systems.
\newblock In {\em FSTTCS'03}, Lecture Notes in Computer Science, pages 73--87,
  2003.

\bibitem[FFHS00]{flamm}
Christoph Flamm, Walter Fontana, Ivo~L. Hofacker, and Peter Schuster.
\newblock {RNA} folding at elementary step resolution.
\newblock {\em RNA}, 6(3):325--338, 2000.

\bibitem[KMM{\etalchar{+}}01]{KMM+}
Yonit Kesten, Oded Maler, Monica Marcus, Amir Pnueli, and Elad Shahar.
\newblock Symbolic model checking with rich assertional languages.
\newblock {\em Theoretical Computer Science}, 256:93--112, 2001.

\bibitem[KSMA03]{kumar-rewriting}
Nirman Kumar, Koushik Sen, José Meseguer, and Gul Agha.
\newblock A rewriting based model for probabilistic distributed object systems.
\newblock In {\em FMOODS'03}, volume 2884 of {\em Lecture Notes in Computer
  Science}, pages 32--46, 2003.

\bibitem[LPRS07]{Lyu07}
Vassily Lyubetsky, Sergey Pirogov, Lev Rubanov, and Alexander Seliverstov.
\newblock Modeling classic attenuation regulation of gene expression in
  bacteria.
\newblock {\em Journal of Bioinformatics and Computational Biology}, 5(1),
  2007.
\newblock in print.

\bibitem[LRSP06]{Lyu06}
Vassily Lyubetsky, Lev Rubanov, Alexander Seliverstov, and Sergey Pirogov.
\newblock Model of gene expression regulation in bacteria via formation of
  {RNA} secondary structures.
\newblock {\em Molecular Biology}, 40(3):440--453, 2006.

\bibitem[RNA]{rnamodel}
RNAmodel.
\newblock Model of {RNA}-related regulation in bacteria.
\newblock http://lab6.iitp.ru/rnamodel/rnamodee.html.

\bibitem[SB91]{SB91}
Maxine Singer and Paul Berg.
\newblock {\em Genes \& genomes}.
\newblock University Science Books Mill Valley, Calif, 1991.

\bibitem[Tou05]{Tou05}
Tayssir Touili.
\newblock Dealing with communication for dynamic multithreaded recursive
  programs.
\newblock In {\em 1st VISSAS workshop}. IOS Press, 2005.

\bibitem[Zuk03]{Zuker03}
Michael Zuker.
\newblock Mfold web server for nucleic acid folding and hybridization
  prediction.
\newblock {\em Nucleic Acids Research}, 31(13):3406--3415, 2003.

\end{thebibliography}

\end{document}